\documentclass[twocolumn,nofootinbib]{revtex4}%

\usepackage{amssymb}
\usepackage{amsfonts}
\usepackage{amsmath}
\usepackage[latin1]{inputenc}
\usepackage{graphicx,color}
\usepackage{slashed}
\usepackage{young}
\usepackage[vcentermath]{youngtab}

\makeatletter

\newcommand{\om}{\omega}

\newcommand{\be}{\begin{equation}}
\newcommand{\ee}{\end{equation}}
\newcommand{\bea}{\begin{eqnarray}}
\newcommand{\eea}{\end{eqnarray}}

\begin{document}

\title{Manifest gravitational duality near anti de Sitter space-time}
\author{Sergio H\"ortner}
\email{sergio.hortner@uam.es}
\affiliation{\medskip
Departamento de F\'isica Te\'orica, Universidad Aut\'onoma de Madrid, 28049 Cantoblanco, Spain}
\affiliation{and}
\affiliation{Erwin Schr\"odinger International Institute for Mathematics and Physics, Boltzmanngasse 9A, 1090 Vienna, Austria}

\begin{abstract}
We derive a manifestly duality-invariant formulation of the Arnowitt-Deser-Misner action principle linearized around anti de Sitter background. The analysis is based on the introduction of two symmetric potentials --on which the duality transformations act-- upon resolution of the linearized constraints, along the lines of previous works focusing on Minkowski and de Sitter backgrounds. Gauge freedom is crucially exploited to solve the constraints in this manner so convenient for exhibiting duality invariance, which suggests a delicate interplay between duality and gauge symmetry. 

\end{abstract}

\maketitle

\section{Introduction}
\setcounter{equation}{0}

The understanding of dualities remains as one of the major challenges of modern theoretical physics. 
Dualities appear in an ample diversity of scenarios --from condensed matter physics to high energy theory--, typically relating strong coupling to perturbative regimes --a rather unique feature that has played a prominent role in the elucidation of non-perturbative aspects of quantum field theory and string theory. In gravitational theories, duality has long been recognized as a constituent of the hidden symmetries that emerge upon toroidal compactifications of eleven-dimensional supergravity \cite{Cremmer} and Einstein gravity \cite{Ehlers}. The rich algebraic structure underlying this phenomenon suggests the existence of an infinite-dimensional Kac-Moody algebra acting as a fundamental symmetry of the uncompactified theory \cite{Julia}-\cite{Lambert} and encompassing the duality symmetries that appear after dimensional reduction. A characteristic property of these algebras is that they involve all the bosonic fields and their Hodge duals, including the graviton and its dual field, and so the associated symmetry transformation for a given tensor field in the bosonic sector relates it to all the rest of the fields (regardless their tensor structure) in a non-trivial way. In four dimensions, the graviton and its dual field are respectively described by symmetric tensors, and it is expected that a duality symmetry --inherited from the underlying infinite-dimensional structure-- relating them may emerge. Naturally, the construction of duality-symmetric action principles constitutes an important part of the program aimed at the investigation of hidden symmetries and dualities in gravity.

\medskip

In this article we show the existence of an off-shell duality symmetry in linearized gravity defined on an anti de Sitter (AdS) background, generalizing previous works where the linearization was performed on Minkowski \cite{HT} and de Sitter (dS) \cite{Julialambda} space-times (see also \cite{Deser} for the case of Maxwell theory). The analysis requires the linearisation of the Arnowitt-Deser-Misner (ADM) action principle \cite{ADM}, \cite{Abbott}, the choice of Poincar\'e coordinates for the AdS background, and the subsequent resolution of the constraints in terms of two symmetric potentials, on which the duality rotations act.

\medskip

The presence of a duality symmetry in the linearized regime near an AdS background was argued in \cite{Julialambda} on the basis of the existence of complex transformations mapping AdS into dS. Concretely, the conformally flat form of de Sitter and anti de Sitter metrics

\bea
d^2_{AdS}&=&\frac{l^2_{AdS}}{r^2}(-dt^2+dr^2+dx^2+dy^2),\nonumber\\
d^2_{dS}&=&\frac{l^2_{dS}}{\eta^2}(-d\eta^2+dx^2+dy^2+dz^2),\nonumber
\eea
are related by the transformation $l_{AdS}^2\rightarrow il_{dS}^2$, $r\rightarrow i\eta$, $t\rightarrow iz$, the time-like boundary of AdS being mapped into a space-like boundary in dS. However, inferring the existence of a duality symmetry in the AdS case from the dS analysis \cite{Julialambda} by this argument implies the isolation of the radial coordinate in the 3+1 space-time splitting. By contrast, our analysis involves the ADM formalism, the isolation of the time-like coordinate and a foliation by space-like hypersurfaces. 

\medskip 

We should also mention that, although our result has not a direct holographic interpretation (for we are dealing with a space-like foliation), the problem of defining duality transformations in gravity linearized around anti de Sitter background has also been addressed from the perspective of holography, motivated by the observation that there is a natural $SL(2,Z)$ action on three-dimensional conformal field theories (CFTs) with $U(1)$ conserved currents, relating the two-point function of the spin-1 conserved current of a given CFT to the two-point function of the spin-1 conserved current of a dual CFT \cite{Witten}. The phenomenon was interpreted as the holographic image of the $SL(2,Z)$ electric-magnetic duality of a $U(1)$ gauge theory defined on the AdS$_4$ bulk. It was subsequently shown that the $SL(2,Z)$ action can be extended to two-point functions of the energy-momentum and higher spin conserved currents in three-dimensional CFTs \cite{petkou}, a result that led the authors to conjecture that linearized higher-spin theories (including spin $s=2$) on AdS$_{4}$ possess a generalization of electric-magnetic duality acting holographically on two-point functions on the boundary. In fact, discrete duality transformations for linearized gravity around AdS with a Pontryagin term --which acts as the analogue of a theta term in electromagnetism-- have been proposed in \cite{petkou} using a time-like slicing of the background geometry. Despite the different character of the space-time splitting employed, it seems appropriate to keep these works in mind when seeking possible extensions of our result that include topological terms.

\medskip

The rest of the article is organized as follows. In Section II we derive the linearization of the ADM action principle around an anti de Sitter background, as well as the form of the gauge transformations of the canonical variables. Section III is dedicated to the resolution of the constraints in terms of potentials. In Section IV we use the expression of the canonical variables in terms of potentials to construct a manifestly duality-invariant action principle. Section V summarizes our results and addresses possible extensions thereof. 

\section{The linearized ADM action principle}
\setcounter{equation}{0}

In order to make manifest the duality symmetry, we shall use the conformal form of the AdS metric (Poincar\'e coordinates):

\be
ds^2=e^{\omega}(dr^2+\eta_{\alpha\beta}dx^{\alpha}dx^{\beta}),\label{ads}
\ee
where $\eta_{\alpha\beta}$ is the three-dimensional Minkowski metric, $\om=log({l}^2/r^2)$ and $l^2=-3/\Lambda$ is the AdS radius.

\medskip

Consider the ADM action principle in the presence of a cosmological constant

\bea
S_{ADM}&=&\int dt d^{3}x[\pi^{ij}\dot{g}_{ij}-N{\cal{H}}-N_{i}{\cal{H}}^{i}].\nonumber\\
\eea
The Hamiltonian and momentum constraints are
\bea
-{\cal{H}}&=&g^{1/2}(^{(3)}R-2\Lambda)+g^{-1/2}(\frac{1}{2}\pi^{2}-\pi^{ik}\pi^{jl}g_{ij}g_{kl}),\nonumber\\
-{\cal{H}}^{i}&=&2\nabla_{j}\pi^{ij},
\eea
and the corresponding Lagrange multipliers are the lapse and shift functions

\be
N=(-g^{00})^{-1/2}, \ \ \ N_{i}=g_{0i}.
\ee
We may perform a power expansion around an AdS background as follows:

\bea
g_{ij}&=&\bar{g}_{ij}+h_{ij}+O(h^2),\nonumber\\
\pi^{ij}&=&\bar{\pi}^{ij}+p^{ij}+O(p^2),\nonumber\\
N_{i}&=&\bar{N}_{i}+h_{0i}+O(h^2),\nonumber\\
N&=&\bar{N}-\frac{1}{2}e^{-\omega/2}h^{00}+O(h^2).\nonumber\\
\eea
The bared quantities correspond to the background space-time, so $\bar{N}_{i}=0$ and $\bar{N}=e^{\om/2}$. The conjugate momentum associated to the background metric is given by

\be
\bar{\pi}^{ij}=\bar{N}\bar{g}^{1/2}\left[\bar{\Gamma}^{0}_{kl}-\bar{g}_{kl}\bar{g}^{mn}\bar{\Gamma}^{0}_{mn}\right]\bar{g}^{ik}\bar{g}^{jl}=-\partial_{0}\omega\delta^{ij},
\ee
and it vanishes in the case of an AdS background. 

\medskip

The linearized action principle reads

\be
S[h_{ij},p^{ij},n,n_{i}]=p^{ij}\dot{h}_{ij}-H-nC-n_{i}C^{i}
\ee
with the Hamiltonian density

\bea
-H&=&\frac{1}{2}e^{\om}p^2-e^{\om}p_{ij}p^{ij}\nonumber\\
&&+\frac{3}{4}e^{-\om}\Delta\om h_{ij}h^{ij}+\frac{1}{2}e^{-\om}h(\partial_{i}\partial_{j}h^{ij}-\Delta h)\nonumber\\
&&+\frac{1}{2}e^{-\om}hh^{ij}\partial_{i}\om\partial_{j}\om+e^{-\om}h\partial_{i}\om\partial^{i}h+e^{-\om}h^{ij}\partial_{i}\partial_{j}h\nonumber\\
&&-\frac{1}{4}e^{-\om}\partial_{i}h_{jk}\partial^{i}h^{jk}-2e^{-\om}\partial_{i}\om h^{ij}\partial_{j}h-\frac{1}{4}e^{-\om}\partial_{i}h\partial^{i}h\nonumber\\
&&+e^{-\om}\partial_{i}h^{ij}\partial_{j}h-e^{-\om}\partial_{j}\om\partial_{i}h^{ij}h\label{H}
\eea
and the constraints
\bea
C&=&e^{-\om}(\partial_{i}\partial_{j}h^{ij}-\Delta h+\partial_{i}\om\partial^{i}h+h\Delta\om), \ \ \\
C^{i}&=&\partial_{j}p^{ij}+\partial_{j}\om p^{ij}-\frac{1}{2}\partial^{i}\om p.\label{constraints}
\eea
These are first-class and generate the gauge transformations

\bea
\delta h_{ij}&=&\partial_{i}\xi_{j}+\partial_{j}\xi_{i}-\xi_{i}\partial_{j}\om-\xi_{j}\partial_{i}\om+\delta_{ij}\partial_{m}\om\xi ^{m}\nonumber\\
&=&e^{\om}[\partial_{i}(e^{-\om}\xi_{j})+\partial_{j}(e^{-\om}\xi_{i})]+\delta_{ij}\partial_{m}\om\xi^{m},\nonumber\\
\delta p_{ij}&=&\delta_{ij}\Delta(e^{-\om}\xi)-\partial_{i}\partial_{j}(e^{-\om}\xi)+\delta_{ij}\partial_{l}\om\partial^{l}(e ^{-\om}\xi).\nonumber\\
\eea
The Lagrange multipliers have been defined as $n_{i}=-2h_{0i}$ and $n=-\frac{1}{2}h^{00}$. The equations of motion for the background metric (see the Appendix) have been used. Indices are raised and lowered with the flat spatial metric $\eta_{ij}$.

\section{Resolution of the constraints}
\setcounter{equation}{0}

We notice that, in order to solve the constraints (\ref{constraints}) in terms of potentials, it is convenient to perform specific gauge transformations that render them in a form similar to the flat background case. Consider the gauge choice

\bea
h_{ij}&=&j_{ij}+e^{\om}[\partial_{i}(e^{-\om}v_{j})+\partial_{j}(e^{-\om}v_{i})]+\delta_{ij}\partial_{m}\om v^{m},\nonumber\\
p^{ij}&=&q^{ij}+\delta^{ij}\Delta u-\partial^{i}\partial^{j}u+\delta^{ij}\partial_{k}\om\partial^{k}u,\label{gaugecond}
\eea
where $j_{ij}$ satisfies $\partial_{i}\omega\partial^{i}j+\Delta\omega j=\partial^{i}(\partial_{i}\omega j)=0$ and $q^{ij}$ is traceless. To prove the existence of such a gauge, it is sufficient to find two particular functions $v_{i}$ and $u$ verifying

\bea
\partial^{i}(\partial_{i}\omega h)=\partial^{i}[\partial_{i}\omega(2\partial_{m}v^{m}+\partial_{m}\om v^{m})]\label{gaugech1}
\eea
and

\bea
p=2\Delta u+3\partial_{m}\omega\partial^{m}u.\label{gaugech2}
\eea
The following choice fulfills the previous requirements:
\bea
v_{i}=\partial_{i}\Delta ^{-1}(e^{\om/2}\frac{h-f(t,x,y)(\partial_{r}\om)^{-1}}{2}),\nonumber\\
u=\frac{e^{-3\om/4}}{2}[\Delta-\frac{15}{8}\Delta\omega]^{-1}[e^{3\om/4}p].
\eea
where $f(t,x,y)$ is a function independent of the radial coordinate $r$, obtained from the integration of (\ref{gaugech1}). In the sequel we shall not specify a particular form for the functions $u$ and $v_{i}$: they will be treated as scalar and vector potentials, respectively. 

\medskip

The constraints now read

\bea
e^{-\om}(\partial_{i}\partial_{j}j^{ij}-\Delta j)&=&0,\label{restc1}\\
e^{-\om}\partial_{j}(e^{\om}q^{ij})&=&0,\label{restc2}
\eea
and remain invariant under the residual gauge transformations
\bea
\delta j_{ij}=\partial_{i}\chi_{j}+\partial_{j}\chi_{i},\label{res1}\\
\delta q^{ij}=e^{-\om}(\delta^{ij}\Delta \chi-\partial^{i}\partial^{j}\chi).\label{res2}
\eea
We may use the residual gauge freedom (\ref{res1}) to carry away the trace of $j_{ij}$. This is clearly consistent with the previous gauge choice (\ref{gaugech1}). The constraint (\ref{restc1}) is then solved in terms of potentials as follows:
\bea
j_{ij}&=&\epsilon_{iab}\partial^{a}\phi^{b}_{\ j}+\epsilon_{jab}\partial^{a}\phi^{b}_{\ i}+\partial_{i}w_{j}+\partial_{j}w_{i},\nonumber\\
\eea
for some vector potential $w_{i}$. On the other hand, the residual gauge freedom (\ref{res2}) may be used to write $q^{ij}$ --constrained to obey $q=0$-- in terms of an unconstrained variable $k^{ij}$ defined as
\bea
q^{ij}=k^{ij}+e^{-\om}(\delta^{ij}\Delta s-\partial^{i}\partial^{j}s)
\eea
for some function $s$ such that $k=-2e^{-\om}\Delta s$. The constraint (\ref{restc2}) is solved as follows:
\bea
q^{ij}&=&e^{-\om}\epsilon^{imn}\epsilon^{jkl}\partial_{m}\partial_{k}P_{nl}+e^{-\om}(\delta^{ij}\Delta s-\partial^{i}\partial^{j}s).\nonumber\\
\eea
An alternative way to derive the previous expression is to first solve (\ref{restc2}) in terms of a constrained potential $Q_{ij}$

\bea
q^{ij}&=&e^{-\om}\epsilon^{imn}\epsilon^{jkl}\partial_{m}\partial_{k}Q_{nl}
\eea
and then write $Q^{ij}=P^{ij}+T^{ij}$ for some unconstrained potential $P^{ij}$ and some tensor $T^{ij}=T^{ij}(P)$ constructed to obey $\epsilon^{imn}\epsilon_{i}^{\ kl}\partial_{m}\partial_{k}P_{nl}=\epsilon^{imn}\epsilon_{i}^{\ kl}\partial_{m}\partial_{k}T_{nl}$ and to generate a gauge transformation of the form (\ref{res2}). The particular choice $T^{ij}=\frac{1}{2}\delta^{ij}(P-\partial_{a}\partial_{b}\Delta^{-1}P^{ab})$ fulfills these conditions.

\medskip

The final expressions for the canonical variables are
\bea
h_{ij}&=&\epsilon_{iab}\partial^{a}\phi^{b}_{\ j}+\epsilon_{jab}\partial^{a}\phi^{b}_{\ i}+\partial_{i}w_{j}+\partial_{j}w_{i}\nonumber\\
&&+e^{\om}[\partial_{i}(e^{-\om}v_{j})+\partial_{j}(e^{-\om}v_{i})]+\delta_{ij}\partial_{m}\om v^{m},\nonumber\\
\\
p^{ij}&=&e^{-\om}\epsilon^{imn}\epsilon^{jkl}\partial_{m}\partial_{k}P_{nl}+e^{-\om}(\delta^{ij}\Delta s-\partial^{i}\partial^{j}s)\nonumber\\
&&+\delta^{ij}\Delta u-\partial^{i}\partial^{j}u+\delta^{ij}\partial_{k}\om\partial^{k}u.\label{hp}
\eea
As observed in the case of Minkowski and de Sitter backgrounds, there is an ambiguity in the definition of the potentials determined by the equations

\bea
\delta h_{ij}&=&\epsilon_{iab}\partial^{a}\delta\phi^{b}_{\ j}+\epsilon_{jab}\partial^{a}\delta\phi^{b}_{\ i}+\partial_{i}\delta w_{j}+\partial_{j}\delta w_{i}\nonumber\\
&&+e^{\om}[\partial_{i}(e^{-\om}\delta v_{j})+\partial_{j}(e^{-\om}\delta v_{i})]+\delta_{ij}\partial_{m}\om\delta v^{m}\nonumber\\
&&=\partial_{i}\xi_{j}+\partial_{j}\xi_{i}-\xi_{i}\partial_{j}\om-\xi_{j}\partial_{i}\om+\delta_{ij}\partial_{m}\om\xi ^{m}\nonumber\\
\eea
and
\bea
\delta p^{ij}&=&e^{-\om}\epsilon^{imn}\epsilon^{jkl}\partial_{m}\partial_{k}\delta P_{nl}+e^{-\om}(\delta^{ij}\Delta\delta s-\partial^{i}\partial^{j}\delta s)\nonumber\\
&&+\delta^{ij}\Delta\delta u-\partial^{i}\partial^{j}\delta u+\delta^{ij}\partial_{k}\om\partial_{k}\delta u\nonumber\\
&=&\delta^{ij}\Delta(e^{-\om}\xi)-\partial^{i}\partial^{j}(e^{-\om}\xi)+\delta^{ij}\partial_{l}\om\partial^{l}(e^{-\om}\xi).\nonumber\\
\eea
They are solved as follows:

\bea
\delta\phi_{ij}&=&\partial_{i}\alpha_{j}+\partial_{j}\alpha_{i}+\delta_{ij}\beta,\nonumber\\
\delta P_{ij}&=&\partial_{i}\gamma_{j}+\partial_{j}\gamma_{i}+\delta_{ij}\eta,\nonumber\\
\delta v_{i}&=&\xi_{i},\nonumber\\
\delta u&=&e^{-\om}\xi,\nonumber\\
\delta w_{i}&=&-\epsilon_{iab}\partial^{a}\alpha^{b},\nonumber\\
\delta s&=&-\eta.
\eea

\section{Manifest duality invariance}
\setcounter{equation}{0}

In this section, we shall use the expression of the canonical variables in terms of the potentials to cast the action principle in a manifestly duality-invariant form. Let us focus first on the kinetic term. Written in terms of the potentials, it reads

\bea
p^{ij}\dot{h}_{ij}=e^{-\om}\epsilon^{imn}\epsilon^{jkl}\epsilon_{iab}\partial_{k}\partial_{m}P_{nl}\partial^{a}\dot{\phi}^{b}_{\ j}.
\eea
The action of the duality transformation $P_{ij}\rightarrow\phi_{ij}$, $\phi_{ij}\rightarrow-P_{ij}$ on the kinetic term yields (up to total derivatives)

\bea
S_{K}\rightarrow S_{K}-\int\;dtd^{3}x\partial_{k}\omega\epsilon^{imn}\epsilon^{jkl}\epsilon_{i}^{\ ab}e^{-\om}\partial_{m}\dot{P}_{nl}\partial_{a}\phi_{bj}.\label{sktransform}\nonumber\\
\eea
The crucial observation is that the extra term in (\ref{sktransform}) can be written as a sum of total derivatives:

\bea
&&-\partial_{k}\omega\epsilon^{imn}\epsilon^{jkl}\epsilon_{i}^{\ ab}e^{-\om}\partial_{m}\dot{P}_{nl}\partial_{a}\phi_{bj}=\nonumber\\
&&\epsilon^{imn}\epsilon^{jkl}\epsilon_{i}^{\ ab}\left\{-\partial_{m}[\partial_{k}\om e^{-\om}\dot{P}_{nl}\partial_{a}\phi_{bj}]\right.\nonumber\\
&&-\partial_{m}[e^{-\om}\partial_{a}\om \dot{P}_{nl}\partial_{k}\phi_{bj}]+\partial_{k}\partial_{a}[e^{-\om}\dot{P}_{nl}\partial_{m}\phi_{bj}]\nonumber\\
&&+\frac{1}{2}\partial_{m}[\partial_{k}\om\partial_{a}\om e^{-\om}\dot{P}_{nl}\phi_{bj}]+\partial_{a}[\partial_{k}\partial_{m}(e^{-\om}\dot{P}_{nl})\phi_{bj}]\nonumber\\
&&-\partial_{a}[\partial_{k}\partial_{m}(e^{-\om}\phi_{bj})\dot{P}_{nl}]+\partial_{k}\partial_{m}[\partial_{a}\om e^{-\om}\dot{P}_{nl}\phi_{bj}]\nonumber\\
&&+\partial_{a}\partial_{m}[\partial_{k}\om e^{-\om}\dot{P}_{nl}\phi_{bj}]-\partial_{k}\partial_{a}[\partial_{m}\om e^{-\om}\dot{P}_{nl}\phi_{bj}]\nonumber\\
&&-\partial_{k}\partial_{m}[e^{-\om}\partial_{a}\dot{P}_{nl}\phi_{bj}]-\partial_{a}\partial_{m}[e^{-\om}\partial_{k}\dot{P}_{nl}\phi_{bj}]\nonumber\\
&&+\left.\partial_{k}\partial_{a}[e^{-\om}\partial_{m}\dot{P}_{nl}\phi_{bj}]\right\}.\label{div}
\eea
Therefore, the kinetic term is invariant under duality transformation (up to total derivatives). The argument can be extended to show the invariance of $S_{K}$ under $SO(2)$ duality rotations (again, up to total derivatives). 

\medskip

On the other hand, substitution of (\ref{hp}) in the Hamiltonian density (\ref{H}) yields:

\bea
-H&=&e^{-\om}[-\epsilon^{imn}\epsilon^{jkl}\partial_{m}\partial_{k}P_{nl}\epsilon_{ipq}\epsilon_{jrs}\partial^{p}\partial^{r}P^{qs}\nonumber\\
&&-\epsilon^{imn}\epsilon^{jkl}\partial_{m}\partial_{k}\phi_{nl}\epsilon_{ipq}\epsilon_{jrs}\partial^{p}\partial^{r}\phi^{qs}\nonumber\\
&&+\frac{1}{2}(\epsilon^{imn}\epsilon_{i}^{\ kl}\partial_{m}\partial_{k}P_{nl})^2+\frac{1}{2}(\epsilon^{imn}\epsilon_{i}^{\ kl}\partial_{m}\partial_{k}\phi_{nl})^2]\nonumber\\
&&-e^{-\om}[\partial_{i}\partial_{j}\phi_{kl}\partial^{i}\partial^{j}\phi^{kl}-\partial_{i}\partial_{j}\phi_{kl}\partial^{k}\partial^{j}\phi^{il}\nonumber\\
&&+\partial^{i}\partial_{j}\phi_{ik}\partial^{j}\partial^{k}\phi-\frac{1}{2}\partial_{i}\partial_{j}\phi\partial^{i}\partial^{j}\phi]\nonumber\\
&&+e^{-\om}[\partial^{i}\partial_{j}\phi_{ik}\partial^{j}\partial_{l}\phi^{kl}-\frac{1}{2}\partial^{i}\partial^{j}\phi_{ki}\partial^{k}\partial^{l}\phi_{jl}]\nonumber\\
&&+3\Delta\om[\partial_{i}\phi_{jk}\partial^{i}\phi^{jk}-\partial_{i}\phi_{jk}\partial^{j}\phi^{ik}-\frac{1}{2}\partial^{j}\phi_{jk}\partial_{i}\phi^{ik}\nonumber\\
&&+\partial^{i}\phi_{ij}\partial^{j}\phi-\frac{1}{2}\partial^{i}\phi\partial_{i}\phi]+\frac{1}{2}e^{-\om}\partial^{i}\om[\partial^{j}\phi_{ik}\partial_{j}\partial_{l}\phi^{kl}\nonumber\\
&&-\partial_{j}\phi_{ik}\partial^{k}\partial_{l}\phi^{jl}+2\partial^{j}\phi_{jk}\partial^{k}\partial^{l}\phi_{il}-2\partial_{j}\phi\partial^{j}\partial^{k}\phi_{ik}\nonumber\\
&&-\partial^{j}\phi_{jk}\partial_{i}\partial_{l}\phi^{lk}+\partial_{j}\phi\partial_{i}\partial_{k}\phi^{jk}+\partial_{j}\phi^{lk}\partial^{j}\partial_{i}\phi_{lk}\nonumber\\
&&-\partial^{j}\phi^{kl}\partial_{i}\partial_{l}\phi_{jk}-\partial_{i}\phi^{jk}\Delta\phi_{jk}+\partial_{j}\phi_{ik}\Delta\phi^{jk}\nonumber\\
&&+3\partial^{j}\phi^{kl}\partial_{j}\partial_{k}\phi_{il}-3\partial^{j}\phi^{kl}\partial_{k}\partial_{l}\phi_{ji}-\partial_{j}\phi^{jk}\Delta\phi_{ik}\nonumber\\
&&+\partial^{j}\phi\Delta\phi_{ij}+\partial_{i}\phi^{jk}\partial_{j}\partial_{k}\phi-\partial_{j}\phi_{ik}\partial^{j}\partial^{k}\phi].
\eea
After integration by parts, the Hamiltonian density can be cast in a more symmetric form:
\bea
-H&=&e^{-\om}[-\epsilon^{imn}\epsilon^{jkl}\partial_{m}\partial_{k}P_{nl}\epsilon_{ipq}\epsilon_{jrs}\partial^{p}\partial^{r}P^{qs}\nonumber\\
&&-\epsilon^{imn}\epsilon^{jkl}\partial_{m}\partial_{k}\phi_{nl}\epsilon_{ipq}\epsilon_{jrs}\partial^{p}\partial^{r}\phi^{qs}\nonumber\\
&&+\frac{1}{2}(\epsilon^{imn}\epsilon_{i}^{\ kl}\partial_{m}\partial_{k}P_{nl})^2\nonumber\\
&&+\frac{1}{2}(\epsilon^{imn}\epsilon_{i}^{\ kl}\partial_{m}\partial_{k}\phi_{nl})^2]+e^{-\om}V
\eea
with

\bea
V&=&3\Delta\om[\partial_{i}\phi_{jk}\partial^{i}\phi^{jk}-\partial_{i}\phi_{jk}\partial^{j}\phi^{ik}-\frac{1}{2}\partial^{j}\phi_{jk}\partial_{i}\phi^{ik}\nonumber\\
&&+\partial_{i}\phi^{ij}\partial_{j}\phi-\frac{1}{2}\partial^{i}\phi\partial_{i}\phi]+\frac{1}{2}\partial^{i}\om[-\partial^{k}\phi_{ji}\partial_{k}\partial_{l}\phi^{jl}\nonumber\\
&&-\partial^{j}\phi_{ik}\partial^{k}\partial^{l}\phi_{jl}+2\partial^{j}\phi_{jk}\partial^{k}\partial^{l}\phi_{li}-2\partial^{k}\phi\partial_{k}\partial^{l}\phi_{il}\nonumber\\
&&+7\partial_{j}\phi^{jk}\partial_{i}\partial^{l}\phi_{kl}-3\partial_{k}\phi\partial_{i}\partial_{j}\phi^{kj}+\partial^{k}\phi_{jl}\partial_{i}\partial_{k}\phi^{jl}\nonumber\\
&&-\partial_{j}\phi_{lk}\partial_{i}\partial^{l}\phi^{jk}+\partial_{i}\phi^{jk}\Delta\phi_{jk}+\partial_{j}\phi_{ik}\Delta\phi^{jk}\nonumber\\
&&+5\partial_{l}\phi^{jk}\partial_{k}\partial^{l}\phi_{ij}-3\partial^{j}\phi^{lk}\partial_{l}\partial_{k}\phi_{ji}-9\partial_{k}\phi^{jk}\Delta\phi_{ji}\nonumber\\
&&+5\partial^{j}\phi\Delta\phi_{ji}+\partial_{i}\phi^{jk}\partial_{j}\partial_{k}\phi-\partial_{j}\phi_{ki}\partial^{j}\partial^{k}\phi\nonumber\\
&&+3\partial^{j}\phi\partial_{i}\partial_{j}\phi-3\partial_{i}\phi\Delta\phi+3\partial_{k}\phi^{jk}\partial_{j}\partial^{l}\phi_{il}\nonumber\\
&&-3\partial^{k}\phi_{ik}\partial_{j}\partial_{l}\phi^{jl}-6\partial_{k}\phi^{jk}\partial_{i}\partial_{j}\phi+6\partial^{j}\phi_{ij}\Delta\phi].\nonumber\\
\eea
One can show that the term $e^{-\om}V$ is a sum of total derivatives, similarly to what we have found in (\ref{div}). The $SO(2)$ duality invariance of the action principle is now manifest.

\section{Conclusions}
\setcounter{equation}{0}

We have shown that linearized gravity around anti de Sitter space-time can be cast in a manifestly duality-invariant form upon resolution of the ADM constraints in terms of two symmetric potentials. The analysis relies on the use of Poincar\'e coordiantes for the AdS background metric. Gauge freedom is exploited in order to introduce the two symmetric potentials in the resolution of the constraints, which suggests a close interplay between duality and gauge symmetry. This result complements previous works where the linearisation was performed around Minkowski and de Sitter space-times, and allows us to conclude that $SO(2)$ duality is a symmetry  of the linearized ADM action around maximally symmetric backgrounds. The structure of the duality-symmetric action principle is similar in the three cases after integrating by parts and dropping boundary terms, the only difference being background-dependent relative factors in the kinetic term and the Hamiltonian. The potentials enjoy the same gauge invariances in the three cases.

\medskip

We have found that duality transformations leave invariant the action principle up to the addition of surface terms. An analogous phenomenon lies at the root of the duality conjecture \cite{petkou} in holography: the introduction of surface terms in the time-like boundary of AdS typically requires the modification of boundary conditions and, since modified boundary conditions are associated with deformations of boundary CFTs, the action of duality in the bulk would imply a transformation of the CFT.

\medskip

An important feature of the potential formalism, which we have also encountered in the present article, is the absence of manifest space-time covariance. Although in some instances it is possible to recover manifest space-time covariance for duality-symmetric action principles (either by the introduction of an infinite number of auxiliary fields \cite{D} with polynomial dependence or a finite number of auxiliary fields with non-polynomial dependence \cite{PST}), when it comes to the case of gravity one may argue that this will probably not be the case by plain contrast of two well-known results. On the one hand, (a discrete version of) electric-magnetic duality is consistent with quantum mechanics \cite{Dirac}. On the other hand, the notion of manifest space-time covariance seems to be inconsistent with the quantum dynamics of gravity \cite{Wheeler}. The immediate conclusion is that, at least in a background-independent approach to quantum gravity, a discrete version of electric-magnetic duality would be allowed, while manifest space-time covariance would not.

\medskip

Last, let us mention possible extensions of the present work. Along the lines of \cite{petkou2}, it would be interesting to consider the inclusion of topological terms in the action principle, in particular the Pontryagin term, then determine whether the constraints are still solvable in terms of potentials and finally search for a (perhaps $SL(2,Z)$) duality-invariant formulation of the action principle. The potential analysis could likewise be performed in the case of a time-like foliation, as a complement to \cite{petkou2}. The derivation of the twisted self-duality equations of motion also deserves investigation, including possible connections with the parent action method for the construction of dual Lagrangians \cite{dual}. The generalisation of our work to the case of arbitrary higher spin fields coupled to a fixed AdS background should as well be studied, building on the works \cite{hs} and \cite{campo}. Finally, it would be interesting to study how the inclusion of boundary counterterms \cite{skendris}-\cite{lee} in AdS affects the potential analysis.

\appendix
\section{}
\subsection*{Einstein equations in a conformally flat background}
\setcounter{equation}{0}

Here we derive Einstein equations for a conformally flat metric and rewrite them in a form particularly convenient for the analysis in the main text. Consider a metric of the form

\be
g_{\mu\nu}=e^{\omega}\eta_{\mu\nu}.
\ee
The associated Riemann tensor, Ricci tensor and scalar curvature are

\bea
R_{\mu\nu\rho\sigma}&=&\frac{1}{2}e^{\om}[\partial_{\mu}\partial_{\sigma}\om\eta_{\nu\rho}+\partial_{\nu}\partial_{\rho}\eta_{\mu\sigma}-\partial_{\nu}\partial_{\sigma}\eta_{\mu\rho}\nonumber\\
&&-\partial_{\mu}\partial_{\rho}\eta_{\nu\sigma}]+\frac{1}{4}e^{\om}[-\partial_{\mu}\om\partial_{\sigma}\om\eta_{\nu\rho}-\partial_{\rho}\om\partial_{\nu}\om\eta_{\mu\sigma}\nonumber\\
&&+\partial_{\mu}\om\partial_{\rho}\om\eta_{\nu\sigma}+\partial_{\nu}\om\partial_{\sigma}\om\eta_{\mu\rho}]\nonumber\\
&&+\frac{1}{4}e^{\om}\partial_{\alpha}\om\partial^{\alpha}\om[\eta_{\mu\sigma}\eta_{\nu\rho}-\eta_{\mu\rho}\eta_{\nu\sigma}],
\eea

\be
R_{\mu\nu}=-\frac{1}{2}[\partial^{\alpha}\partial_{\alpha}\om+\partial^{\alpha}\om\partial_{\alpha}\om]\eta_{\mu\nu}-\partial_{\mu}\partial_{\nu}\om+\frac{1}{2}\partial_{\mu}\om\partial_{\nu}\om,
\ee

\be
R=3e^{-\om}[-\partial^{\alpha}\partial_{\alpha}\om-\frac{1}{2}\partial^{\alpha}\om\partial_{\alpha}\om].
\ee
The Einstein equations

\be
R_{\mu\nu}-\frac{1}{2}Rg_{\mu\nu}+\Lambda g_{\mu\nu}=0
\ee
imply
\be
R_{\mu\nu}=\Lambda g_{\mu\nu}\label{Einstein}
\ee
after taking the trace $R=4\Lambda$. The components of (\ref{Einstein}) are

\be
R_{0i}=\partial_{0}\partial_{i}\om-\frac{1}{2}\partial_{0}\om\partial_{i}\om=0,
\ee
\be
R_{00}=-\frac{3}{2}\partial_{0}\partial_{0}\om+\frac{1}{2}\Delta\om+\frac{1}{2}\partial_{i}\om\partial^{i}\om=-\Lambda e^{\om},\label{00}
\ee
and
\bea
R_{ij}&=&-\partial_{i}\partial_{j}\om+\frac{1}{2}\partial_{i}\om\partial_{j}\om\nonumber\\
&&-\frac{1}{2}\delta_{ij}(-\partial_{0}\partial_{0}\om+\Delta\om-\partial_{0}\om\partial_{0}\om+\partial_{i}\om\partial^{i}\om)=e^{\om}\delta_{ij}\Lambda.\nonumber\\\label{3}
\eea
Using the trace of (\ref{3}) we derive the equation
\be
-\partial_{i}\partial_{j}\om+\frac{1}{2}\partial_{i}\om\partial_{j}\om-\frac{1}{2}\delta_{ij}(-\frac{2}{3}\Delta\om+\frac{1}{3}\partial_{i}\om\partial^{i}\om)=0.\label{tr}
\ee
From (\ref{00}) and (\ref{tr}) we get

\be
\partial_{0}\partial_{0}\om-\frac{1}{2}\partial_{0}\om\partial_{0}\om=-\frac{1}{3}\Delta\om+\frac{1}{6}\partial_{i}\om\partial^{i}\om.
\ee
The conformally flat form of the anti de Sitter metric (\ref{ads}) verifies $\partial_{0}\omega=0$. Using this condition we find

\be
\Delta\omega-\frac{1}{2}\partial_{i}\omega\partial^{i}\omega=0,
\ee
\be
\Delta\om=-\frac{2}{3}\Lambda e^{\om},
\ee
so
\be
\partial_{i}\partial_{j}\omega-\frac{1}{2}\partial_{i}\omega\partial_{j}\omega=0\label{ij}.
\ee
Equation (\ref{ij}) and the condition $\partial_{0}\om=0$ contain all the information of Einstein equations for the AdS background, and are thoroughly used in the main text.

\section*{Acknowledgments} 
It is a pleasure to thank Enrique Alvarez, Jos\'e  Barb\'on, Andrea Campoleoni, Thomas Curtright, Stefan Fredenhagen, Luis Ib\'a\~nez, Bernard Julia and Tom\'as Ort\'in for their valuable support. This research work received funding from the Spanish Research Agency (Agencia Estatal de Investigacion) through the grant IFT Centro de Excelencia Severo Ochoa SEV-2016-0597 and from the Erwin Schr\"odinger International Institute for Mathematics and Physics through a Junior Research Fellowship..

\smallskip

This article is dedicated to the memory of Rosario ``Charo'' Aranda.


\begin{thebibliography}{99}

\bibitem{Cremmer}
  E.~Cremmer and B.~Julia,
  ``The SO(8) Supergravity,''
  Nucl.\ Phys.\ B {\bf 159}, 141 (1979).

\bibitem{Ehlers}

J. Ehlers, ``Transformation of static exterior solutions of Einstein's gravitational field equations into different solutions by means of conformal mappings,'' in  ``Les Theories relativistes de la gravitation'', Colloques Internationaux du CNRS 91, 275 (1962); R.~P.~Geroch, ``A Method for generating solutions of Einstein's equations,'' J.\ Math.\ Phys.\  {\bf 12}, 918 (1971); R.~P.~Geroch,  ``A Method for generating new solutions of Einstein's equations. 2,'' J.\ Math.\ Phys.\  {\bf 13}, 394 (1972).
  

\bibitem{Julia}
B.~Julia, ``Group disintegrations,'' in ``Superspace and Supergravity'', Hawking, S.W., and Ro\u{c}ek, M., eds., Nuffield Gravity Workshop, Cambridge, England, June 22 - July 12, 1980 (Cambridge University Press, Cambridge, U.K.; New York, U.S.A., 1981) p. 331-350; B.~L.~Julia, ``Dualities in the classical supergravity limits: Dualizations, dualities and a detour via (4k+2)-dimensions,'' in *Strings, branes and dualities. Proceedings, NATO Advanced Study Institute, Cargese, France, May 26-June 14, 1997* Ed. L.~Baulieu, P.~Di Francesco, M.~Douglas, V.~Kazakov, M.~Picco, P.~Windey, NATO Sci.Ser.C 520 (1999)
Dordrecht, Netherlands, p. 121-139.

\bibitem{Nicolai}
T.~Damour, M.~Henneaux and H.~Nicolai, ``E10 and a ``small tension expansion'' of M
Theory'', Phys. Rev. Lett. 89 221601 (2002).

\bibitem{West}
  P.~C.~West,
  ``E(11) and M theory,''
  Class.\ Quant.\ Grav.\  {\bf 18}, 4443 (2001).

\bibitem{Lambert}
N.~D.~Lambert, P.~C.~West, 
``Coset Symmetries in Dimensionally Reduced Bosonic String Theory,''
Nucl.\ Phys. \ B {\bf 615}, 117 (2001).
  
\bibitem{HT}
  M.~Henneaux and C.~Teitelboim,
  ``Duality in linearized gravity,''
  Phys.\ Rev.\  D {\bf 71}, 024018 (2005).

	\bibitem{Julialambda} 
  B.~Julia, J.~Levie and S.~Ray,
  ``Gravitational duality near de Sitter space,''
  JHEP {\bf 0511}, 025 (2005).

\bibitem{Deser}
S.~Deser and C.~Teitelboim,
``Duality transformations of Abelian and non-Abelian gauge fields,''
Phys.\ Rev.\ D 13 1592  (1976).

\bibitem{ADM}
R.~Arnowitt, S.~Deser and C.~W.~Misner,
``The Dynamics of General Relativity,'' in  	\textsl{Gravitation: an introduction to current research}, Louis Witten ed. Wiley (1962).

\bibitem{Abbott}
L.~F.~Abbott and S.~Deser,
``Stabilty of gravity with a cosmological constant,''
Nucl.Phys. B195 (1982).

\bibitem{D}
B. McClain, Y.S. Wu, and F. Yu, Nucl. Phys.B343 (1990) 689; C. Wotzasek, Phys. Rev. Letters 66 (1991) 129
F.P. Devecchi and M. Henneaux, Phys. Rev. D54 (1996) 1606; I. Bengtsson and A. Kleppe, .
I. Martin and A. Restuccia, Phys. Lett. B323 (1994) 311.
N. Berkovits, Phys. Lett. B388 (1996) 743


\bibitem{PST}
P.~Pasti, D.~Sorokin and M.~Tonin, Phys.~Rev.~D55, 6292 (1997); P.~Pasti, D.~Sorokin and M.~Tonin, Phys.~Rev.~D52 (1995) 4277.

\bibitem{Dirac}
P.~A.~M.~Dirac, ``Quantised Singularities in the Electromagnetic Field,'' Proc.~Roy.~Soc.~A 133, 60 (1931); P.~A.~M.~Dirac, ``The theory of magnetic poles,'' Phys. Rev. 74, 817 (1948).

\bibitem{Wheeler}
J.~A.~Wheeler, in ``Battelle Rencontres: 1967 Lectures on Mathematical Physics''
(Benjamin, New York, 1968); B.~S.~DeWitt, Phys. Rev. 160 (1967) 1113.


\bibitem{Witten}
E.~Witten, ``SL(2,Z) action on three-dimensional conformal field theories with Abelian symmetry,'' In Shifman, M. (ed.) et al.: From fields to strings, vol. 2 1173-1200.


\bibitem{petkou}
R.~G.~Leigh, A.~C.~Petkou, ``SL(2,Z) Action on Three-Dimensional CFTs and Holography,'' JHEP 0312 (2003) 020; R.~G.~Leigh.
 
\bibitem{petkou2}
A.~C.~Petkou, ``Gravitational duality transformations on (A)dS$_4$,'' JHEP 0711 079 (2007).


\bibitem{dual}
N.~Boulanger, A.~Campoleoni, I.~Cortese, L.~Traina, ``Spin-2 twisted duality in (A)dS,'' Front.in Phys. 6 (2018) 129; N.~Boulanger, A.~Campoleoni, I.~Cortese, ``Dual actions for massless, partially-massless and massive gravitons in (A)dS,'' Phys.Lett. B782 (2018) 285-290. 


\bibitem{hs}
M.~Henneaux, S.~H\"ortner, A.~Leonard, ``Higher Spin Conformal Geometry in Three Dimensions and Prepotentials for Higher Spin Gauge Fields,'' JHEP 1601 (2016) 073.


\bibitem{campo}
A.~Campoleoni, M.~Henneaux, S.~H\"ortner, A.~Leonard, ``Higher-spin charges in Hamiltonian form. I. Bose fields,'' JHEP 1610 (2016) 146; A.~Campoleoni, M.~Henneaux, S.~H\"ortner, A.~Leonard, ``Higher-spin charges in Hamiltonian form. II. Fermi fields,'' JHEP 1702 (2017) 058.

\bibitem{skendris}

M.~Henningson, K.~Skenderis, ``The holographic Weyl anomaly,'' JHEP07, 023 (1998).

\bibitem{stresstensor}
V.~Balasubramanian, P.~Kraus, ``A stress tensor for Anti de Sitter gravity,'' Commun.Math.Phys. 208 (1999).

\bibitem{lee}
S.~Hyun, W.T.~Kim, J.~Lee, ``Statistical entropy and AdS/CFT correspondence in BTZ black holes,'' Phys. Rev.D59 084020 (1999).

\end{thebibliography}
\end{document}